\documentclass{emulateapj}
\usepackage{apjfonts}
\usepackage[]{natbib}
\usepackage{graphics}
     
\newcommand{\etal}{et al.}  
\newcommand{\per}{\ensuremath{^{-1}}}
\newcommand{\persq}{\ensuremath{^{-2}}}

\newcommand{\hal}{H\ensuremath{\alpha}}
\newcommand{\hbeta}{H\ensuremath{\beta}} 
\newcommand{\hst}{\emph{HST}}
\newcommand{\msun}{\ensuremath{M_{\odot}}}
 
\newcommand{\kms}{km~s\ensuremath{^{-1}}} 
\newcommand{\ebv}{\ensuremath{E(B-V)}}

\newcommand{\mbh}{\ensuremath{M_\mathrm{BH}}}

\newcommand{\sigmastar}{\ensuremath{\sigma_\star}}

\newcommand{\msigma}{\ensuremath{\mbh-\sigma}}

\newcommand{\lbol}{\ensuremath{L_{\mathrm{bol}}}}

\newcommand{\ledd}{\ensuremath{L_{\mathrm{Edd}}}}

\newcommand{\mstar}{\ensuremath{M_\star}}

\shorttitle{NUCLEAR CLUSTER AND BLACK HOLE IN NGC 3621}
\shortauthors{BARTH ET AL.}

\begin{document} 

\title{Dynamical Constraints on The Masses of the Nuclear Star Cluster
  and Black Hole in the Late-Type Spiral Galaxy NGC 3621}

\author{Aaron J. Barth\altaffilmark{1}, Louis E.
  Strigari\altaffilmark{1}, Misty C. Bentz\altaffilmark{1}, Jenny
  E. Greene\altaffilmark{2,3}, and Luis C. Ho\altaffilmark{4}}

\altaffiltext{1}{Department of Physics and Astronomy, 4129 Frederick
  Reines Hall, University of California, Irvine, CA 92697-4575;
  barth@uci.edu}

\altaffiltext{2}{Department of Astrophysical Sciences, Princeton
  University, Princeton, NJ 08544}

\altaffiltext{3}{Hubble Fellow and Princeton-Carnegie Fellow}

\altaffiltext{4}{The Observatories of the Carnegie Institution of
  Washington, 813 Santa Barbara Street, Pasadena, CA 91101}

\begin{abstract}

NGC 3621 is a late-type (Sd) spiral galaxy with an active nucleus,
previously detected through mid-infrared [\ion{Ne}{5}] line emission.
Archival \emph{Hubble Space Telescope} (\hst) images reveal that the
galaxy contains a bright and compact nuclear star cluster.  We present
a new high-resolution optical spectrum of this nuclear cluster,
obtained with the ESI Spectrograph at the Keck Observatory.  The
nucleus has a Seyfert 2 emission-line spectrum at optical wavelengths,
supporting the hypothesis that a black hole is present.  The
line-of-sight stellar velocity dispersion of the cluster is
$\sigmastar=43\pm3$ \kms, one of the largest dispersions measured for
any nuclear cluster in a late-type spiral galaxy.  Combining this
measurement with structural parameters measured from archival \hst\
images, we carry out dynamical modeling based on the Jeans equation
for a spherical star cluster containing a central point mass.  The
maximum black hole mass consistent with the measured stellar velocity
dispersion is $3\times10^6$ \msun.  If the black hole mass is small
compared with the cluster's stellar mass, then the dynamical models
imply a total stellar mass of $\sim1\times10^7$ \msun, which is
consistent with rough estimates of the stellar mass based on
photometric measurements from \hst\ images.  From structural
decomposition of 2MASS images, we find no clear evidence for a bulge
in NGC 3621; the galaxy contains at most a very faint and
inconspicuous pseudobulge component ($M_K\gtrsim-17.6$ mag).  NGC 3621
provides one of the best demonstrations that very late-type spirals
can host both active nuclei and nuclear star clusters, and that
low-mass black holes can occur in disk galaxies even in the absence of
a substantial bulge.

\end{abstract}

\keywords{galaxies: active --- galaxies: individual (NGC 3621) ---
  galaxies: kinematics and dynamics --- galaxies: nuclei --- galaxies:
  spiral}

\section{Introduction}

In the study of the demographics of supermassive black holes, one
important unresolved problem is the observational census of black
holes in very late-type disk galaxies.  The masses of supermassive
black holes in elliptical and early-type spiral galaxies appear to be
well correlated with the bulge properties of their host galaxies
\citep{kr95,fm00,geb00}.  These correlations naturally lead to the
question of whether supermassive black holes can form in disk galaxies
that lack bulges, and if so, how the black hole mass in bulgeless
galaxies might be related to the properties of the host.  The nearest
example of a bulgeless disk galaxy, M33, does not contain a massive
black hole, with an extremely tight upper limit of $\mbh \lesssim
1500-3000$ \msun\ determined from stellar-dynamical observations and
modeling \citep{geb01, mfj01}.  It is not yet known whether the
absence of a massive black hole is a typical property of bulgeless
spirals in general, however, because very few late-type disk galaxies
are near enough for stellar-dynamical measurements to yield such
stringent constraints on the black hole mass.

It has long been recognized that many spiral galaxies contain
photometrically and dynamically distinct central star-cluster nuclei,
with the M33 nucleus being one of the best-studied examples
\citep[e.g.,][]{wal64,vdb76,ggm82,na82,oco83}.  The \emph{Hubble Space
Telescope} (\hst) made it possible to survey the properties of nuclear
star clusters in more distant galaxies \citep[e.g.,][]{phil96, car97},
and \hst\ imaging programs have demonstrated that the majority of very
late-type spirals (Hubble types Scd--Sd) contain compact star-cluster
nuclei at or very close to their isophotal centers \citep{mat99,
bok02}.  Spectroscopic observations have revealed that the nuclear
clusters in late-type spirals typically have stellar velocity
dispersions of $\sim15-35$ \kms\ \citep{bok99,wal05}, effective radii
of a few parsecs \citep{bok04}, and dynamical masses of
$\sim10^6-10^7$ \msun\ \citep{bok99,wal05}.  The spectra also reveal
composite stellar populations indicative of multiple star formation
episodes, with most clusters containing a population component with an
age of $\lesssim100$ Myr \citep{wal06,ros06}.

The discovery of scaling relationships between nuclear cluster masses
and host galaxy properties, both in spirals \citep{ros06} and in dwarf
ellipticals \citep{wh06,fer06}, offers an intriguing hint of a
connection between the processes that govern the growth of nuclear
clusters and black holes.  \citet{fer06} speculate that the formation
of nuclear star clusters and massive black holes might be mutually
exclusive, such that black holes are able to form only in galaxies
above some critical mass.  However, the only nuclear star clusters in
which the presence of a massive black hole can be ruled out at any
significant level are in the Local Group or within a few Mpc.  Optical
spectroscopic surveys have shown that active galactic nuclei (AGNs) do
occur in some nuclear star clusters in nearby galaxies
\citep{seth08a}, so black holes and nuclear clusters can apparently
coexist.  Evidence that black holes can occur in at least some very
late-type disk galaxies comes from the detection of a small number of
AGNs in Scd and Sd-type spirals.  The best example is the Sd galaxy
NGC 4395, which contains a Seyfert 1 nucleus \citep{fs89, fh03}; it
remains the only clear identification of a broad-lined AGN in a
bulgeless disk galaxy.  In addition, a few examples of Type 2 AGNs in
very late-type spirals have been detected recently in optical
spectroscopic surveys, such as NGC 1042 \citep{seth08a,shi08} and UGC
6192 \citep{bar08}.  In both NGC 4395 and NGC 1042 the AGNs occur in
nuclear star clusters (it is not known whether UGC 6192 contains a
nuclear cluster).

A recent \emph{Spitzer} spectroscopic observation of the Sd galaxy NGC
3621 by \citet{sat07} led to the discovery of an active nucleus, based
on the detection of [\ion{Ne}{5}] emission lines at 14.3 \micron\ and
24.3 \micron.  Since photon energies greater than 95 eV are required
for photoionization of Ne$^{+3}$ to Ne$^{+4}$, ordinary \ion{H}{2}
regions are not expected to be significant sources of [\ion{Ne}{5}]
emission, but a hard AGN continuum can easily provide the necessary
ionizing photons.  Emission lines from a range of ionization states of
neon are observable in the mid-infrared, and the relative strengths of
these lines are useful as diagnostics of the ionization conditions
within AGN narrow-line regions \citep[e.g.,][]{sm92,v92,stu02,gds06}.
\citet{as08} used the results of new photoionization models to argue
that the strength of the [\ion{Ne}{5}] emission in NGC 3621 can only
be plausibly explained by the presence of an AGN, and not by an
ordinary burst of nuclear star formation.  The detection of an AGN in
NGC 3621 is of significant interest because it is one of the few very
late-type spirals known to host an active nucleus, making it an
important target for further observations to constrain its black hole
mass and AGN energetics.  \citet{sat07} note that there is no
previously published optical spectrum of the nucleus of NGC 3621
suitable for emission-line classification.

In this paper, we use archival \hst\ images to show that NGC 3621
contains a well-defined and compact nuclear star cluster.  A new
optical spectrum of this star cluster is used to examine the
classification of the active nucleus, and to measure the stellar
velocity dispersion of the cluster.  We describe dynamical modeling of
the nuclear cluster and the resulting constraints on the masses of
both the cluster and the central black hole.  We also examine the
structure of NGC 3621 using near-infrared images from 2MASS in order
to search for a bulge component in this late-type galaxy.  For the
distance to NGC 3621, we adopt $D=6.6$ Mpc, based on the Cepheid
measurements of \citet{fre01}.  At this distance, 1\arcsec\
corresponds to 32.0 pc.

\section{Imaging Data}

\subsection{Archival HST Data}

NGC 3621 has been observed numerous times with \hst, although many of
the observations were of outer fields that do not include the nucleus,
taken as part of the \hst\ Key Project on the Cepheid distance scale
\citep{fre01}.  We found three sets of images that did cover the
galaxy nucleus: one with WFPC2, one with ACS, and one with NICMOS.
The parameters for each of these observations are listed in Table
\ref{hsttable}.  The galaxy was observed with ACS/WFC at two separate
pointings.  One ACS pointing placed the nucleus on the WFC1 CCD and
the other pointing placed it on the WFC2; at both positions, identical
exposure sequences were obtained in three filters.  We selected the
WFC1 pointing since this image included a larger region surrounding
the nucleus.  The NICMOS F190N exposure was taken as the continuum
observation for an F187N (Paschen $\alpha$) narrow-band image.  We do
not use the F187N emission-line image here.

The NICMOS and ACS images were retrieved from the \hst\ data archives
and we use the standard pipeline-processed versions of these images.
For the WFPC2 data, we use the cosmic-ray cleaned and co-added image
from the WFPC2 Associations archive.

\begin{figure}
\plotone{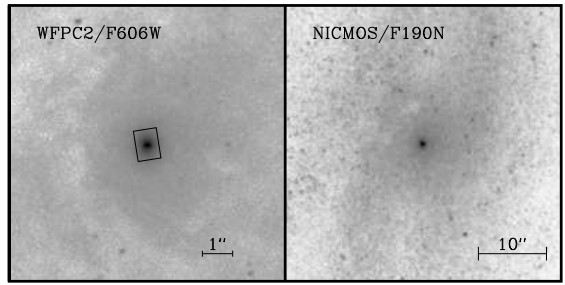}
\caption{\hst\ images of the nucleus of NGC 3621.  \emph{Left panel:}
  A portion of the WFPC2/PC F606W image.  The rectangle surrounding
  the nucleus shows the position and size of the spectroscopic
  aperture used in the Keck observation, for the 1\arcsec-wide
  extraction described in \S\ref{sectionreductions}.  \emph{Right
  panel:} A portion of the NICMOS/NIC3 F190N image, with a larger
  field of view displayed.  In both panels, north is up and east is to
  the left.
\label{hstimages}}
\end{figure}

Figure \ref{hstimages} displays the central regions of the WFPC2 and
NICMOS images.  The images show that the galaxy contains a very
compact and photometrically distinct nuclear star cluster.
Surrounding the cluster is a smooth and nearly featureless region of
radius $\sim1\farcs5$ (or $\sim50$ pc). At larger radii, dust lanes
and young star clusters become more prominent in the optical images,
especially in the ACS F435W band.

\begin{deluxetable}{lccc}
\tablewidth{3.3in} \tablecaption{Hubble Space Telescope Archival Data}
\tablehead{\colhead{Camera} & \colhead{Filter} & \colhead{Exposure
Time (s)} & \colhead{Observation Date} } 
\startdata 
WFPC2/PC & F606W &
$2\times80$ & 1994-08-17 \\ 
ACS/WFC1 & F435W & $3\times360$ & 2003-02-03 \\
 & F555W & $3\times360$ & 2003-02-03 \\
 & F814W &
$3\times360$ & 2003-02-03 \\ 
NICMOS/NIC3 & F190N & $6\times224$ & 2007-03-31
\enddata
\label{hsttable}
\end{deluxetable}

\subsubsection{GALFIT Modeling}
\label{wfpc2galfit}

To determine the structure of the nuclear cluster, we use the
2-dimensional modeling package GALFIT \citep{peng02}.  Unfortunately,
the nuclear star cluster is saturated in the ACS/WFC F555W and F814W
images, making it impossible to use these images to derive structural
parameters for the cluster.  The ACS F435W image is not saturated, but
in this blue passband the dust lanes and massive stars in the
circumnuclear region are more prominent, making this band less easily
suited to modeling.  In the NICMOS image (with a scale of 0\farcs2
pixel\per), the cluster is unresolved.  Therefore, we use the WFPC2/PC
image for the GALFIT decomposition.  WFPC2 magnitudes listed below are
on the Vegamag system, using a zeropoint of 22.887 for the F606W
filter \citep{bag97}.

For the GALFIT modeling, we extracted a $101\times101$ pixel$^2$
region from the PC image, centered on the nuclear cluster.  This
corresponds roughly to the smooth region surrounding the cluster.  A
larger region would be less suitable for GALFIT modeling because of
the increasing prominence of the dust lanes at larger distances from
the nucleus.  A few patches or lanes of dust are visible at the edges
of this extracted image, and we created a mask image to exclude these
regions from the GALFIT optimization.  For the point-spread function
(PSF) model, we used the Tiny Tim package \citep{kri93} to create a
PSF for the WFPC2/PC camera and F606W filter, at $2\times$
oversampling.  In GALFIT, the galaxy model is created on an
oversampled grid, convolved with the PSF model, and finally resampled
back to the WFPC2/PC plate scale and convolved with the WFPC2 CCD
charge-diffusion kernel \citep{kri93} in order to compare with the
observed galaxy image.

As an initial step, we attempted to model the image as the sum of an
exponential disk plus a central cluster, using a variety of possible
models for the cluster based on the various functional forms available
in GALFIT (Gaussian, Moffat function, S\'ersic function, and King
model).  None could adequately represent the cluster profile; in each
case there were large systematic residuals.  After much
experimentation we took a different approach, modeling the cluster as
a sum of several Gaussian components, since a superposition of a
sufficient number of Gaussians can in principle be used to fit an
arbitrary cluster profile.  As a simplifying step, we constrained the
Gaussian components and exponential disk to be concentric in the
GALFIT model, and found that we were still able to fit the cluster
profile adequately with this restriction.  We began with an
exponential disk plus a single Gaussian component and added additional
Gaussians until the model profile ceased to improve significantly with
the addition of a further component.  In our final model we use five
Gaussian components plus an exponential disk.

\begin{figure}
\plotone{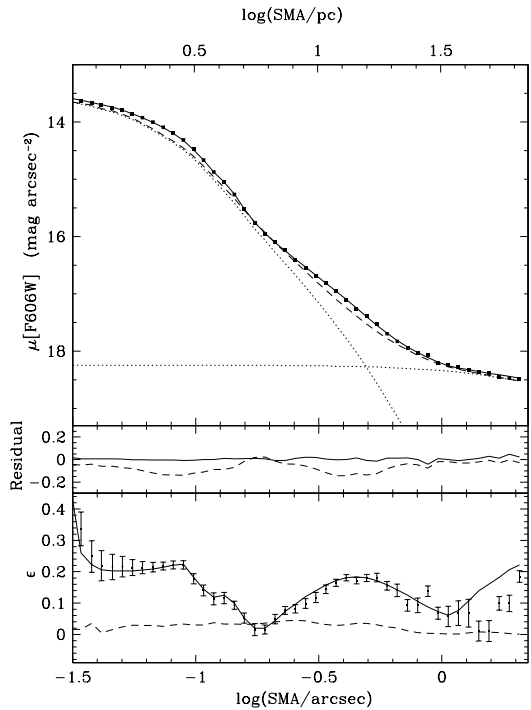}
\caption{Radial profiles of surface brightness and ellipticity for the
  nuclear cluster in the WFPC2/PC F606W image.  ``SMA'' denotes the
  semi-major axis of the fitted elliptical isophotes.  The individual
  data points represent the radial profile measured from the
  \emph{HST} image.  The long-dashed curve is the radial profile of
  the circularly symmetric GALFIT model, and the solid curve is the
  radial profile of the model fit that was not constrained to have
  circular symmetry.  Dotted curves show the radial profiles for the
  disk and cluster separately, for the symmetric model.  The middle
  panel shows the model residuals for the radial profile fit (in mag
  arcsec\persq).  The slight departures from zero ellipticity in the
  circularly symmetric model are due to random errors in the ellipse
  fitting to the GALFIT model.
\label{wfpc2radprof}}
\end{figure}

We created two versions of this GALFIT model.  In the first, the
individual Gaussian components were forced to be concentric and
circularly symmetric (i.e., axis ratio $b/a=1$).  This model lends
itself most easily to deprojection and dynamical modeling, but might
not yield an acceptable fit to the cluster structure.  In the second
version, the individual components were still set to be concentric,
but we allowed each Gaussian component to have arbitrary values of
axis ratio and position angle.

We used the IRAF\footnote{IRAF is distributed by the National Optical
Astronomy Observatories, which are operated by the Association of
Universities for Research in Astronomy, Inc., under cooperative
agreement with the National Science Foundation.}  \texttt{ellipse}
task to examine the radial profile of the cluster and GALFIT models;
this task uses the elliptical isophote fitting method described by
\citet{jed87}.  Figure \ref{wfpc2radprof} illustrates the results.  We
find that the two models (symmetric and asymmetric) have nearly
identical radial profiles.  The asymmetric model yields an excellent
match to the cluster's ellipticity profile (with ellipticity ranging
from $\sim0.015$ to 0.38 within the inner $r<2\arcsec$).  Although the
symmetric model is obviously unable to match the ellipticity profile
of the cluster, the azimuthally-averaged surface brightness profile of
the symmetric model is very similar to that of the cluster itself, as
seen in Figure \ref{wfpc2radprof}.  This makes it possible to use a
deprojected version of the symmetric model as a simplified,
spherically-symmetric representation of the cluster structure (see
\S\ref{dynamicalmodeling}).

In the spherical model, the individual Gaussian components have FWHMs
of 1.8, 4.3, 9.9, 20.9, and 69.7 pixels (with an image scale of
0\farcs045 pixel\per).  The three narrowest components contribute the
majority of the cluster's light.  The broadest Gaussian component from
the GALFIT decomposition has an extent (FWHM=69.7 pixels or 3\farcs1)
much larger than the visible size of the cluster in the WFPC2 image,
and we conclude that it represents light belonging predominantly to
the galaxy disk rather than to the cluster.  As we show below in
\S\ref{dynamicalmodeling}, the choice of whether to assign this
component to the cluster or the disk has a very small effect on the
dynamical modeling results.  The next-broadest component, with
FWHM=0\farcs94, has a size comparable to the wings of the cluster
visible in the WFPC2 image (as well as being similar in size to our
spectroscopic aperture).  We interpret this component as belonging to
the cluster, although the distinction between cluster and disk becomes
ambiguous in the outer wings of this component and there is no clear
way to determine the outermost extent of the cluster.

The F606W magnitudes of these five Gaussian components are 17.86,
18.29, 17.97, 18.09, and 17.18 mag, respectively. If we represent the
cluster by the sum of the first four Gaussian components, then its
half-light radius is 0\farcs13, corresponding to 4.2 pc.
\citet{bok04} find that 50\% of nuclear clusters in late-type spirals
have $r_e$ between 2.4 and 5.0 pc, so the size of the NGC 3621 nuclear
cluster is very typical for this class of objects.  The exponential
component in the GALFIT decomposition has a scale length of 23\arcsec\
and a magnitude of $m_\mathrm{F606W} = 9.7$ mag, corresponding to
$V\approx10.0$ mag based on an estimated color of
$(\mathrm{F606W}-V)=-0.3$ mag for an Sc galaxy template \citep{kc96}.
Since the GALFIT fitting was carried out over only the central
$4\farcs5\times4\farcs5$ of the galaxy, there is no reason to expect
this component to be an accurate representation of the primary
exponential disk of the entire galaxy, but its magnitude is
surprisingly close to the galaxy's total $V$ magnitude of 9.5 mag as
listed in the HyperLeda catalog \citep{pat03}.

The radial profile plot shown in Figure \ref{wfpc2radprof} shows that
the observed (i.e., PSF-convolved) surface brightness falls by $\sim5$
mag arcsec\persq\ within the inner $r=1$ arcsec of the galaxy profile,
where the cluster outskirts have essentially merged into the
surrounding disk.  Such a large gradient in surface brightness over
such a small radius places this object among the most solidly detected
nuclear clusters, in comparison with the late-type spirals from the
\citet{bok02} sample.

\subsubsection{Aperture Photometry}

Unfortunately, the saturation in the ACS F555W and F814W images
precludes their use for GALFIT modeling or aperture photometry of the
nuclear cluster.  If the default ACS/WFC \texttt{GAIN=2} setting had
been used for these observations, it would still have been possible to
perform accurate aperture photometry on mildly saturated objects.
With \texttt{GAIN=2}, saturation of the CCD full well results in
charge-bleeding onto adjacent pixels while still conserving the total
countrate for the object \citep{gil04}.  However, the NGC 3621 images
were taken with \texttt{GAIN=1}, which does not sample the CCD full
well capacity, and in this situation the photometry of saturated
objects is compromised.  This leaves the ACS F435W, WFPC2 F606W, and
NICMOS F190N observations as the only usable \hst\ images.  The NICMOS
image has a much coarser plate scale (0\farcs2 pixel\per) and a much
broader PSF than the optical images, and the impact of dust lanes is
very different between the optical and near-infrared images, and as a
result it would be very difficult to carry out GALFIT modeling in a
consistent manner for all of these images.  Therefore, we fall back on
a simpler strategy, in which we perform aperture photometry to measure
the cluster's brightness in each image.

Under the assumption that the narrowest four Gaussian components from
the GALFIT model represent the light of the cluster, 90\% of the
cluster's light (in projection) is contained within a circular
aperture of radius 12 pixels, or 0\farcs54.  We use this as the
fiducial aperture radius for the photometric measurements, since at
larger radii it becomes very difficult to distinguish between the
outskirts of the cluster and the surrounding galaxy disk.  For the
photometry, we set the background annulus close to the cluster, in
order to minimize uncertainties due to the gradient in the fairly flat
light profile of the host galaxy, and we use a background region of
$0\farcs6<r<1\farcs0$.  For the two optical images, we find \hst\
filter magnitudes for the cluster of $m_\mathrm{F435W} = 17.82$ and
$m_\mathrm{F606W} = 16.75$ mag (both given on the Vega system).
Uncertainties on these measurements are dominated by the choice of the
inner and outer background radii (rather than photon-counting
statistics), and we estimate overall uncertainties of $\sim0.1$ mag on
each measurement based on varying the inner and outer radii of the
background region over wider radial ranges.  Using the IRAF
\texttt{SYNPHOT} package, we can convert these \hst\ filter magnitudes
to Johnson $B$ and $V$ magnitudes.  For a wide range of spectral
shapes corresponding to S0--Sc galaxy templates \citep{kc96}, the
(F606W$-V$) color index is $-0.32\pm0.02$ mag.  The ACS F435W filter
is very close to a Johnson $B$ passband, and \texttt{SYNPHOT}
calculates a color index of (F435W$-B$)$=0.03\pm0.01$ mag for the
\citet{kc96} templates.  Correcting for Galactic extinction based on
$E(B-V)=0.08$ mag \citep{sfd98}, we have $M_B=-11.65$ and $M_V =
-12.30$ mag for the cluster, not including any correction for internal
extinction within the host galaxy or the cluster itself.

We carried out photometry using the same aperture and background radii
on the NICMOS image, but an additional aperture correction was
required in order to use photometry through an aperture of the same
physical size, because the NICMOS NIC3 PSF contains significant flux
at radii larger than 0\farcs54.  To estimate the aperture correction,
we created a GALFIT model of the nuclear cluster at the plate scale of
the NIC3 camera, using the components from the WFPC2 GALFIT model, and
convolved the model with a NIC3 PSF generated by Tiny Tim.  We then
carried out aperture photometry on the simulated cluster image in an
0\farcs54 aperture, and determined the aperture correction needed to
correct that photometric measurement to one that would correspond to
90\% of the total cluster light (to be consistent with the optical
measurements).  The resulting aperture correction was 0.23 mag.
Applying the NICMOS photometric zeropoint to the aperture-corrected
photometric measurement, we find a flux density in the F190N filter of
$f_\nu=2.4\times10^{-26}$ ergs cm\persq\ s\per\ Hz\per\ at 19000 \AA,
corresponding to an AB magnitude of 15.43 or a Vega magnitude of 13.83
mag.

\subsection{2MASS Data}

To determine the overall structure of the galaxy, we examine images
from the 2MASS survey \citep{skr06}.  Of the three 2MASS bands, the
\emph{H}-band image of NGC 3621 has the best $S/N$ and image quality,
and we used this image to create the primary GALFIT model.  The image
scale is 1\arcsec\ pixel\per, corresponding to 32 pc pixel\per.  While
the nuclear cluster is unresolved in the 2MASS images, it is still
visible as a distinct, compact peak in the galaxy center.

We created a PSF model for the 2MASS image using the \texttt{PSF} task
in the IRAF DAOPHOT package, combining the profiles of 42 individual
isolated stars from the image.  The resulting PSF model is well fit by
a Moffat function with FWHM = 2\farcs89.

\begin{figure}
\plotone{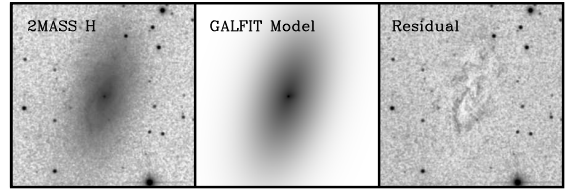}
\caption{\emph{Left panel:} 2MASS \emph{H}-band image of NGC 3621.
  \emph{Center panel:} GALFIT model.  \emph{Right panel:} GALFIT
  residuals.  The image size is $5\arcmin\times5\arcmin$.  North is up
  and east is to the left.
\label{2massimages}}
\end{figure}

\begin{figure}
\plotone{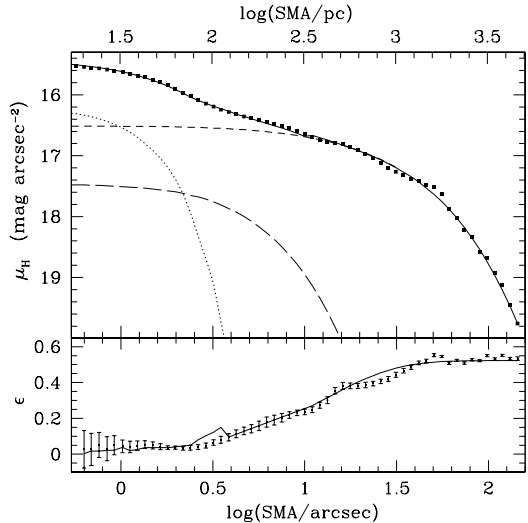}
\caption{Radial surface brightness profile (upper panel) and
  ellipticity profile (lower panel) measured from the 2MASS
  \emph{H}-band image using the IRAF \texttt{ellipse} task.  The
  individual points show the profile of the 2MASS image, and the solid
  curve is the full galaxy model.  Individual model profiles are shown
  for the nuclear cluster (dotted line), inner exponential
  (long-dashed line), and outer exponential component (short-dashed
  line).  The nuclear cluster is modeled as an unresolved source and
  this component illustrates the radial profile of the PSF image.
\label{2massradprof}}
\end{figure}

To construct a GALFIT model of the galaxy, we started with a single
exponential disk component and added additional components as needed
to eliminate strong systematic residuals.  A three-component model
(convolved with the PSF) proved to be the minimal model that yielded
an acceptable fit, with the three components describing the primary
disk, an inner disk or possible bulge component, and the nuclear
cluster.  We used a S\'ersic model for the possible bulge component.
With the S\'ersic index $n$ left to vary freely in the fit, both $n$
and $r_e$ for this component converged to extremely large and
unreasonable values.  To better constrain the model, we carried out
the fit with the S\'ersic index $n$ of the inner component fixed to
values 1, 2, 3, and 4, and the model with $n=1$ (i.e., an exponential
profile) yielded the lowest $\chi^2$.  We modeled the nuclear cluster
as an unresolved source and found an acceptable fit with no strong
residuals at the location of the nucleus.  The 2MASS image,
best-fitting model, and fit residuals are shown in Figure
\ref{2massimages}, and the galaxy radial profile is shown in Figure
\ref{2massradprof}.  Magnitudes for each model component are listed in
Table \ref{2masstable}.  For the outer exponential disk, we find a
scale length of $h=45\farcs6$ (or 1.46 kpc) and axis ratio of
$b/a=0.48$.  For the inner S\'ersic component with $n=1$, the model
yields $r_e=9\farcs6$ (or 307 pc), and $b/a=0.64$.

Following the same procedures, we created PSF models for the 2MASS $J$
and $K$-band images, and ran GALFIT on the $J$ and $K$ images using
the best-fitting $H$-band model parameters as an initial input to the
fit.  The fits were very similar to the $H$-band results, and the
component magnitudes are listed in Table \ref{2masstable}.  As is
typically the case for GALFIT decompositions, the formal fitting
errors returned by GALFIT are extremely small ($<0.02$ mag) and only
reflect the statistical uncertainties on the best-fit model.  The
actual uncertainties on each component magnitude are dominated by
systematic errors, typically driven by real structural differences
between the galaxy's light profile and the idealized GALFIT model
(such as overall asymmetries, dust lanes, and spiral structure), and
are probably at least 0.1 mag.

\begin{deluxetable}{lccc}
\tablewidth{3.3in} \tablecaption{2MASS GALFIT Fitting Results}
\tablehead{\colhead{Component} & \colhead{\emph{J}} &
  \colhead{\emph{H}} & \colhead{\emph{K}} } 
\startdata 
Primary Disk           & 7.61  & 6.96  & 6.71 \\
Inner S\'ersic ($n=1$) & 12.65 & 11.73 & 11.47 \\ 
Nucleus                & 14.14 & 13.53 & 13.32 
\enddata
\tablecomments{All results are given in magnitudes, on the 2MASS
  photometric system.}
\label{2masstable}
\end{deluxetable}

The primary exponential disk is the dominant component of the galaxy,
containing 99\% of the total $H$-band light in the galaxy model.  Our
fitting results demonstrate that the galaxy does not contain any
substantial bulge component, and possibly no bulge at all.  At best
there is a minor plateau in the inner disk surface brightness,
corresponding to the inner S\'ersic component in the GALFIT fits, that
might possibly be ascribed to a faint pseudobulge.  The central
surface brightness of this inner exponential component is 1 mag
arcsec\persq\ fainter than the surface brightness of the primary
(outer) disk component.
 
The inner S\'ersic component does have some properties in common with
pseudobulges.  It is best fit with an exponential profile ($n=1$),
which is typical for pseudobulges in spiral galaxies
\citep{as94,apb95}.  Furthermore, the ratio of its effective radius to
the scale length of the primary disk is $r_e/h = 0.21$, consistent
with the mean $\langle r_e/h\rangle = 0.22\pm0.09$ found by
\citet{mch03} for a large sample of late-type spirals containing
pseudobulges.  With an absolute magnitude of $M_H=-17.4 , M_K=-17.6$
mag, the luminosity of the inner exponential component is within the
range spanned by pseudobulges in nearby disk galaxies \citep{car01} as
well.  The inner S\'ersic component also has a smaller ellipticity
than the primary disk, suggestive of a bulge-like feature; this is
also seen as the gradual increase in ellipticity with radius in the
$H$-band radial profile (Figure \ref{2massradprof}).  Nevertheless,
this component is so faint relative to the primary disk that it is
uncertain whether it represents a distinct physical component of the
galaxy.  Since its interpretation is somewhat uncertain, we take its
luminosity to be an upper limit to the luminosity of a bulge or
pseudobulge in NGC 3621.  Deeper and higher-resolution infrared
observations would still be desirable, in order to carry out a more
definitive search for a faint bulge component in this galaxy.

One possible concern in these model fits is that a small bulge, nearly
unresolved in the 2MASS images, might be subsumed within the nuclear
cluster component in the model fits.  However, the small size of this
component makes this unlikely, because the majority of the light in
the unresolved nuclear component is contained within the inner $r=2$
pixels (64 pc).  Still, to test for the presence of non-cluster light
in the nuclear GALFIT component, we compared the 2MASS fitting results
to the small-aperture \hst\ NICMOS F190N measurement of the nuclear
cluster brightness.  The NICMOS aperture photometry results in
$f_\nu=2.4\times10^{-26}$ ergs cm\persq\ s\per\ Hz\per\ at 19000 \AA.
Converting the nuclear cluster magnitudes from the 2MASS GALFIT models
($H=13.52, K=13.32$) to flux densities, we find $f_\nu$(1.65
\micron)$=4.0\times10^{-26}$ and $f_\nu$(2.2
\micron)$=3.0\times10^{-26}$ ergs cm\persq\ s\per\ Hz\per, in each
case somewhat larger than the NICMOS flux.  This suggests that there
may be some contribution of non-cluster light included in the nuclear
cluster component in the GALFIT models, which would not be surprising
given the coarse pixel scale of the 2MASS images.  Still, the amount
of possible ``extra'' non-cluster light in the nuclear cluster model
component is well below the brightness of the inner exponential
component in the GALFIT models, and the upper limits to the bulge
luminosity described above are not affected.

\subsection{Stellar Mass Estimates}
\label{sectionstellarmass}

The stellar mass of the cluster is a quantity of primary interest.
Ideally, the stellar mass would be determined from \hst\ photometry
using the widest possible wavelength coverage, from the ultraviolet to
near-infrared, as well as stellar population fits to spectroscopic
data, in order to minimize degeneracies between age, metallicity, and
reddening.  In addition, the stellar population is expected to be
composite, containing a wide range of stellar ages, as seen in other
nuclear clusters.  The accuracy of our estimates is very limited with
only three photometric bands available.  The lack of photometric
coverage blueward of the $B$ band exacerbates the problem, since
$U$-band or near-UV measurements would be needed to accurately
constrain the presence of a young stellar population component in the
cluster.  Furthermore, the available spectroscopic data (described
below) has very low $S/N$ blueward of \hbeta, making it unsuitable for
the sort of detailed stellar population modeling that was done by
\citet{wal06} and \citet{ros06} for other nuclear clusters.  

To estimate the cluster's stellar mass, we used the \texttt{kcorrect}
package (version 1.4; Blanton \& Roweis 2007), which performs fits of
model spectral energy distributions for composite stellar populations
with varying amounts of reddening to photometric data in arbitrary
filter sets.  The code uses a basis set of instantaneous-burst spectra
from the models of \citet{bc03} with the \citet{chab03} initial mass
function, and finds the best-fitting match of the spectral templates
to the photometric data using a nonnegative matrix factorization
technique.  To input our photometric data points into
\texttt{kcorrect}, we correct them for Galactic extinction ($A=0.33$,
0.22, and 0.04 mag in the F435W, F606W, and F190N filters,
respectively), and convert to AB magnitudes.  The code returns a
best-fitting mass-to-light ratio of 0.47 (in solar units) in the
NICMOS F190N band, and a stellar mass of $\mstar=1.5\times10^7$ \msun\
based on the cluster light within the 0\farcs54 photometric aperture.
Combining the mass estimate from \texttt{kcorrect} with the cluster's
$V$-band luminosity, we have a $V$-band mass-to-light ratio of
$(M/L)_V = 2 (\msun/L_{\odot,V})$.

To obtain an additional estimate of the stellar mass from the
photometric measurements, we applied the results of \citet{bell03} to
derive a $K$-band mass-to-light ratio based on the optical color
measured from the ACS and WFPC2 images.  The \citet{bell03} $M/L$
values are based on the PEGASE spectrophotometric evolution models
\citep{fr97}, assuming an exponentially declining star-formation rate.
The assumed stellar initial mass function (IMF) is a modified Salpeter
(1955) IMF with a smaller contribution from low-mass stars, resulting
in a 30\% reduction in mass relative to a Salpeter IMF for a given
luminosity.  Bell \etal\ estimate that their $M/L$ ratios may have
$\sim20\%$ random uncertainties and $\sim25\%$ systematic
uncertainties due to dust extinction and differences in galaxy ages
and star-formation histories.  Uncertainties in the IMF are likely to
contribute at least this much systematic error as well.

Transforming the optical photometry to Johnson filters, we find
$(B-V)\approx0.6$ mag for the cluster.  For this color index, the Bell
\etal\ results predict $(M/L)_K = 0.75$ in solar units.  Applying this
mass-to-light ratio to the point source component in the 2MASS
$K$-band model gives a stellar mass of $3\times10^7$ \msun.  As
discussed above, the comparison between NICMOS and 2MASS photometry of
the nuclear cluster suggests that the 2MASS $K$ magnitude might be too
bright by $\sim20\%$ due to inclusion of non-cluster light.  The Bell
\etal\ prescriptions yield a $V$-band stellar mass-to-light ratio of
$(M/L)_V = 1.4 (\msun/L_{\odot,V})$, and combining this with the
absolute $V$ magnitude determined from the WFPC2 data implies $\mstar
= 1.0\times10^7$ \msun.  From all of these results, our best estimate
of the cluster's stellar mass is $\sim(1-3)\times10^7$ \msun.

These stellar masses are highly uncertain due to the limited spectral
coverage of the \hst\ data.  Even with a more complete dataset,
however, there are potentially large systematic uncertainties in
stellar masses obtained by these techniques.  As discussed by
\citet{kg07}, different methods for estimating galaxy stellar masses
(i.e., from spectral energy distribution model fitting with different
evolutionary synthesis models, or from the Bell \etal\ $M/L$
calibration) can result in systematic differences of at least a factor
of 2, and possibly more than a factor of 3, even when the same IMF is
assumed.  Allowing for uncertainty in the IMF, a systematic
uncertainty of more than a factor of 4 in stellar mass is certainly
plausible, particularly in view of the complex star formation history
of a typical nuclear star cluster.  Ultimately, the main conclusion
that we can draw from these results is that within the substantial
uncertainties, the stellar mass derived from the \hst\ photometry
appears to be consistent with the cluster's dynamical mass (as
described below in \S\ref{dynamicalmodeling}).

\section{Spectroscopic Data}

\subsection{Observations and Reductions}
\label{sectionreductions}

A 600 s exposure of the nucleus of NGC 3621 was obtained at the
Keck-II telescope on the night of 2008 March 2 with the Echellette
Spectrograph and Imager \citep[ESI;][]{she02}.  An 0\farcs75-wide slit
was used in ESI echelle mode, and the slit was oriented at the
parallactic angle (PA=9\arcdeg).  The instrumental dispersion with
this slit width is $\sigma_i \approx 22$ \kms, and the pixel scale in
the dispersion direction is 11.5 \kms\ pixel\per, with total
wavelength coverage of 3850--11000 \AA\ over 10 echelle orders.  We do
not have a direct measurement of the seeing during the exposure (taken
at airmass 1.62), but the seeing in standard star observations on that
night was $\sim0\farcs8-1\farcs2$.  Extracted spectra were
wavelength-calibrated using observations of HgNe, Xe, and CuAr
comparison lamps, and flux-calibrated with an observation of the
standard star Feige 34.  Error spectra were extracted and propagated
through the same sequence of calibrations.

We performed the spectral extractions twice using different extraction
widths.  To obtain a spectrum of the nuclear cluster, we extracted the
spectrum using an extraction width of 1\arcsec, and a background
region of 2\farcs5--5\farcs0 on either side of the nucleus.  This
small-aperture spectrum did not yield satisfactory results for the
emission lines, however.  The emission lines are extended over a
region of several arcseconds along the slit, and the emission-line
surface brightness is not centrally concentrated.  Consequently, the
small-aperture extraction included only a small fraction of the total
emission-line flux in the slit, and the close-in background regions
had nearly the same emission-line surface brightness as the extraction
window.  In the resulting spectrum, the emission lines had erroneously
low fluxes due to this background subtraction problem.  To obtain a
better emission-line spectrum, we performed another extraction using a
5\arcsec\ extraction width, with the background regions close to the
edges of the slit where the emission-line surface brightness is lower.
We use the narrow extraction in \S\ref{section:kinematics} below for
measurement of the nuclear cluster's velocity dispersion, and the
wider extraction in \S\ref{section:emissionlines} to examine the
nuclear emission-line ratios.

\subsection{Stellar Kinematics}
\label{section:kinematics}

\begin{figure}
\plotone{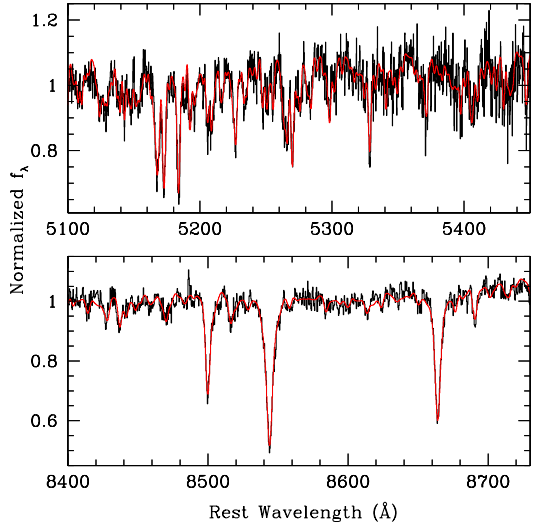}
\caption{Spectral regions used for measuring the stellar velocity
  dispersion.  The red curve is the best-fitting velocity-broadened
  template star spectrum.
\label{veldisp}}
\end{figure}

The velocity dispersion of the nuclear star cluster was measured with
the direct-fitting routine described by \citet{bhs02}, using spectra
of four stars of spectral type G8III--K2III observed during the same
run.  We performed fits separately to spectral regions around Mg$b$
(5100-5450 \AA) and the \ion{Ca}{2} triplet lines (CaT; 8400-8750
\AA).  The fitting routine models the galaxy spectrum as the sum of a
stellar spectrum broadened by a Gaussian kernel and a power-law
featureless continuum.  The velocity-broadened spectrum is multiplied
by a quadratic polynomial that can account for both reddening and
minor differences in flux calibration between the galaxy and template
star. The resulting spectrum is then fitted to the data, with the
velocity dispersion, power-law slope and normalization, and
multiplicative polynomial parameters allowed to vary as free
parameters in the fit.  The measurements for the blue and red spectral
regions yielded identical results of $\sigma=43\pm3$ \kms, where the
uncertainty represents the sum in quadrature of the fitting error on
the best-fitting template and the standard deviation of the velocity
dispersions derived from all template stars.  Figure \ref{veldisp}
shows the template fits to the spectrum of the nuclear cluster.  The
Gaussian-broadened template spectrum fits the profiles of the strong
\ion{Ca}{2} lines well, indicating that the possible contribution of
higher-order moments to the line-of-sight velocity profile is
relatively insignificant for the integrated spectrum of the cluster.

For the blue spectral region, differences in the the Fe/$\alpha$
abundance ratio between the galaxy and the template stars can
potentially lead to biased results, because the strongest features
immediately redward of Mg$b$ are Fe features such as Fe5270.  The
direct-fitting routine requires the featureless continuum dilution to
vary smoothly with wavelength, and it cannot properly account for
strong differences in line strength between adjacent Fe and Mg
features (see Barth \etal\ 2002 for further details).  However, in
this case we found that the fitting results were not sensitive to
whether the Mg$b$ lines were included or masked out in the fit, and
Figure \ref{veldisp} shows that the routine yields a good fit over the
entire 5100--5450 \AA\ region.  As an additional check, we also
measured the velocity dispersion using an independent direct-fitting
code described by \citet{gh06a} and found very consistent results
($44\pm2$ and $43\pm2$ \kms\ for the Mg$b$ and CaT regions,
respectively).  We adopt $\sigmastar=43\pm3$ \kms\ as the best
estimate of the line-of-sight stellar velocity dispersion.

The ESI slit length of 20\arcsec\ is too short to obtain a pure sky
spectrum simultaneously with the galaxy observation, and it is
unavoidable that the background regions used for sky subtraction
contain light from the galaxy disk.  Consequently, the background
subtraction employed in the spectral extraction must be subtracting
off some galaxy light with a different velocity dispersion and
possibly a different mean velocity than that of the cluster itself.
We tested the magnitude of this effect using the data from the ESI
echelle order containing Mg$b$ and Fe5270.  For this order, we added
the extracted sky spectrum back into the extracted cluster spectrum,
to obtain a spectrum of the cluster without sky subtraction.  We then
ran the velocity dispersion fitting routine on this spectrum.  The
result was an increase of only 1 \kms\ in the measured dispersion, in
comparison with the value measured from the sky-subtracted spectrum.
Thus, we conclude that the sky subtraction is not adversely affecting
the velocity dispersion measurement.  In the spectral order containing
the \ion{Ca}{2} triplet lines, the background spectrum is dominated by
night sky emission lines, making it impossible to carry out a similar
test for the \ion{Ca}{2} dispersion.

Even when background subtraction is applied during the spectral
extraction, the spectrum includes some residual light from the
underlying galaxy disk.  We used the results of the GALFIT
decomposition of the WFPC2 image to estimate the fractional
contribution of galaxy disk light to the ESI spectrum in the $V$ band.
We took the GALFIT models for the nuclear cluster and disk separately,
convolved each with a Gaussian kernel of FWHM = 1\farcs2 to represent
the likely seeing during the Keck observation, and measured the flux
from each component within an aperture corresponding to the ESI slit
($0\farcs75\times1\arcsec$).  Within this aperture, the disk
contributes 26\% of the total light.  However, the background
subtraction used in the spectroscopic extraction reduces this disk
contamination substantially.  The background regions used were
2\farcs5-5\farcs0 on either side of the nucleus.  We used the PC image
itself to estimate the flux level in the background regions, since the
GALFIT model does not include real structure such as dust lanes that
are present in the actual sky background regions.  We convolved the PC
F606W image with the same Gaussian kernel and measured the counts in
the exact background regions used in the ESI extraction.  After
subtraction of the mean background level, the remaining contribution
of galaxy disk light to the ESI spectrum is only 10\%.  From the
results of the background-subtraction test described above, we
conclude that this small residual contamination of disk light in the
ESI spectrum should not have a significant effect on the measured
velocity dispersion of the cluster.

It is important to note that our measured value of $\sigmastar$ is
essentially the second moment of the cluster's integrated
line-of-sight velocity profile, which may contain a contribution from
rotation as well as from random stellar motions within the cluster.
\citet{seth08b} have found from high-resolution integral-field
spectroscopic data that the nuclear cluster in NGC 4244 (an edge-on
Scd galaxy) is rotating at 30 \kms, and if nuclear clusters are formed
\emph{in situ} by inflowing gas in the host galaxy disk, rapid
rotation might be an important feature of nuclear clusters in general.

\subsection{Emission-Line Spectrum}
\label{section:emissionlines}

\begin{figure}
\plotone{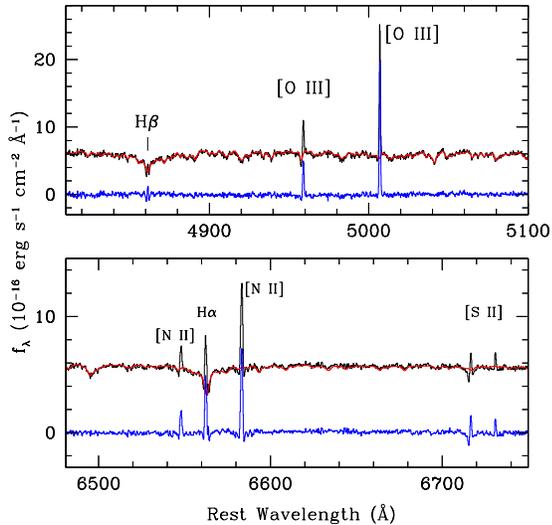}
\caption{Starlight subtraction in the spectral region surrounding the
  \hbeta\ and [\ion{O}{3}] lines (upper panel) and the \hal\ and
  [\ion{N}{2}] lines (lower panel).  In each panel the spectrum of the
  nuclear cluster is shown in black, the best-fitting starlight model
  in red, and the starlight-subtracted spectrum in blue.
\label{starsub}}
\end{figure}

To measure the emission-line strengths and ratios, we carried out
starlight subtraction on the wide-aperture spectral extraction using
similar techniques, this time using a linear combination of a
G8III--K3III giant and an A0V star spectrum to represent the stellar
continuum.  The starlight-subtraction fits were carried out over the
ranges 4600--5300 \AA\ and 6000--7000 \AA, with the wavelengths of the
narrow emission lines masked out in the fit.  The results are shown in
Figure \ref{starsub}.  The starlight subtraction works reasonably well
but results in an imperfect fit to the \hal\ and \hbeta\ absorption
lines, leaving some spurious residual absorption in the
starlight-subtracted spectrum.  This residual absorption is the main
source of uncertainty in the emission-line flux measurements described
below, particularly for the very weak \hbeta\ line.

\begin{figure}
\plotone{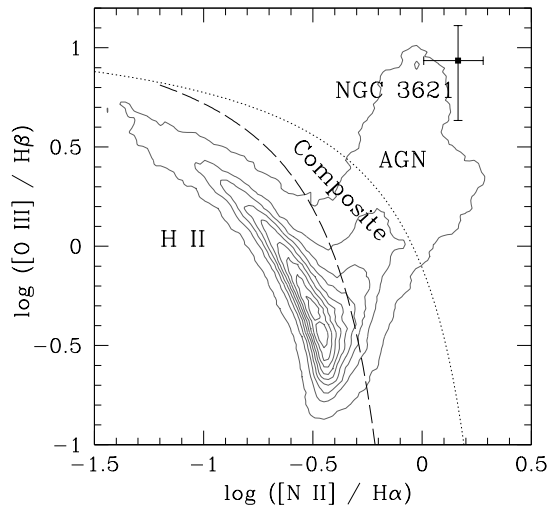}
\caption{Emission-line diagnostic diagram illustrating the AGN
  classification of the NGC 3621 nucleus.  Contours represent SDSS
  emission-line galaxies from \citet{kau03}.  The dashed and dotted
  lines represent the classification boundaries of \citet{kew06}.
\label{bpt}}
\end{figure}

Emission line fluxes were measured by direct integration of the
starlight-subtracted spectrum, with the results listed in Table
\ref{linefluxes}.  For \hal\ and \hbeta, we attempted to correct for
the starlight-subtraction residuals by manually fitting a spline to
the residual absorption profile and subtracting it from the spectrum;
we performed several trial subtractions and the uncertainty range in
the measured fluxes reflects the range of values measured in different
trials.  Although some line fluxes are highly uncertain, the results
provide a clear emission-line classification for the nucleus.  Figure
\ref{bpt} shows the location of the NGC 3621 nucleus in a line-ratio
diagnostic diagram, in which it falls in the main Seyfert branch of
the diagram.  Thus, we classify NGC 3621 as a Seyfert 2 galaxy, and
the optical spectrum adds further evidence that this galaxy contains
an AGN.  

From inspection of the two-dimensional ESI spectrum, the emission-line
properties change systematically at a distance of roughly 3\farcs5
from the nucleus.  At this radius, the [\ion{O}{3}]/\hal\ and
[\ion{N}{2}]/\hal\ ratios appear to drop substantially.  If this
change in emission-line ratios marks the boundary of the AGN-dominated
narrow-line region, then the radius of this region is $\sim110$ pc.
Assuming a circularly symmetric narrow-line region on the plane of the
sky, the ESI spectrum includes only $\sim10\%$ of the total area of
the emission-line region.  Within this region, the emission-line
surface brightness appears nearly constant in the ESI spectrum, thus
our spectrum probably contains only about 10\% of the total
narrow-line flux of the AGN.  It would be very useful to obtain a
nuclear spectrum with an integral-field spectrograph, in order to
better map out this emission-line region and obtain improved estimates
of the total AGN luminosity.

\begin{deluxetable}{lc}
\tablewidth{2.5in} \tablecaption{Emission-Line Measurements}
\tablehead{\colhead{Line} & \colhead{Flux} }
\startdata
\hbeta\ & $2.6\pm1.3$ \\
{}[\ion{O}{3}] $\lambda5007$ & $22.4\pm0.4$ \\
\hal\ & $8.9\pm2.4$ \\
{}[\ion{N}{2}] $\lambda6583$ & $13.0\pm1.8$ \\
{}[\ion{S}{2}] $\lambda6716$ & $2.0\pm0.5$ \\ 
{}[\ion{S}{2}] $\lambda6731$ & $1.8\pm0.5$ 
\enddata
\tablecomments{Emission-line fluxes are given in units of $10^{-16}$
  ergs cm\persq\ s\per, and are corrected for Galactic reddening of
  $E(B-V)=0.08$ mag \citep{sfd98}.  The flux scale has an
  unknown overall scaling uncertainty due to slit losses.}
\label{linefluxes}
\end{deluxetable}

The emission-line widths are unresolved in the ESI spectrum.  We find
FWHM $\approx52$ \kms\ for arc lamp lines observed through the same
0\farcs75 slit width.  The raw linewidths (not corrected for
instrumental broadening) are 51 \kms\ for [\ion{O}{3}] $\lambda5007$
and 57 \kms\ for [\ion{N}{2}] $\lambda6583$, so the intrinsic
linewidths must be substantially smaller than the instrumental
resolution.  Such narrow lines are exceptionally unusual for Seyfert 2
nuclei, even in low-mass host galaxies.  \citet{bar08} used this same
ESI setup to measure the linewidths for a sample of low-mass Seyfert 2
galaxies selected from the Sloan Digital Sky Survey, and the smallest
[\ion{O}{3}] width found was FWHM=66 \kms\ in the late-type spiral UGC
6192.  NGC 3621 is so nearby that the ESI spectrum subtends only a
small portion of the narrow-line region, and the linewidth might be
substantially different if measured through a larger aperture.

The internal reddening toward the narrow-line region cannot be
determined accurately, because of the very uncertain \hbeta\ flux
measurement.  From the Balmer line fluxes (corrected for Galactic
extinction) we find \hal/\hbeta$=3.4\pm1.9$.  This implies a possible
range for the internal reddening from $\ebv=0$ up to $\ebv=0.6$ mag,
assuming Case B recombination for the intrinsic Balmer-line ratios.

\section{Dynamical Modeling}
\label{dynamicalmodeling}

As an initial approximation of the cluster's total mass, we use
the virial theorem for a spherical, isotropic, and isolated system:
\begin{equation}
M_\mathrm{vir} = f\frac{r_e \sigmastar^2}{G}.
\label{eq:virial}
\end{equation}
Here, $f$ is a dimensionless factor that depends on the density
profile and velocity dispersion profile of the cluster, and
$\sigmastar$ is the line-of-sight velocity dispersion averaged over
the density profile of the cluster.  The numerical value of $f$ is
typically $\sim10$ \citep[e.g.,][]{wal05} if the effective radius
$r_e$ is used as the cluster radius parameter.  Taking $r_e\approx4$
pc and $\sigmastar=43$ \kms, this yields $M_\mathrm{vir} \approx
1.7\times10^7$ \msun.

Though the virial theorem provides a rough estimate of the cluster
mass, the precise value of the factor $f$ depends on the density
profile and velocity dispersion profile of the cluster.  Additionally,
the value of $f$ would be affected if there are multiple components
that significantly contribute to the mass of the cluster, such as a
point mass in the form of a central black hole.  In this case, the
outskirts of the cluster blend smoothly into the galaxy disk, and the
outer radius of the cluster is not well determined.  For these
reasons, in order to derive a more accurate cluster mass, as well as
to determine constraints on the mass of the black hole, we compute
dynamical models for the cluster based on the Jeans equation. This
formalism allows us to model the structure of a star cluster with a
two-component mass distribution including stars and a central point
mass.

Similar to the virial theorem method, the total cluster mass as
derived from the Jeans equation depends on its outer extent.  However,
determining the extent of the light that is gravitationally bound to
the cluster is somewhat complicated by the fact that it is embedded in
a disk.  Particularly near the outermost radii, it is not entirely
straightforward to distinguish between the light from stars bound to
the cluster and the light due to the stars in the disk.  In the GALFIT
model for the cluster, 90\% of the total cluster light in the
multi-Gaussian model is contained within a projected radius of
0\farcs54, or $\sim17$ pc.  To conservatively model the mass bound to
the cluster, we use $r_\mathrm{out}=17$ pc as the fiducial outer
radius of the cluster.  All results described below from the dynamical
modeling refer to the mass enclosed within this radius.  We also note
that the results described here refer to the mass of the nuclear
cluster; the mass contributed by the galaxy disk component within this
fiducial radius is not included in the dynamical mass results quoted
below.  Within a projected radius of 17 pc, the total light from the
four GALFIT components corresponding to the cluster is equivalent to
$L = 7.8\times10^6 ~L_{\sun,V}$.

The Jeans equation for the radial velocity dispersion, $\sigma_r$, is 
 \begin{equation}
r \frac{d(\rho_{\star} \sigma_r^2)}{dr} =  - \rho_{\star}(r) 
\frac{GM(r)}{r}- 2 \beta(r) \rho_{\star}(r) \sigma_r^2.
\label{eq:jeans}
\end{equation}
The anisotropy of the stellar distribution is defined as $\beta = 1 -
\sigma_\theta^2/\sigma_r^2$, the total dynamical mass enclosed within
radius $r$ is $M(r)$, and the three-dimensional luminosity density is
$\rho_\star(r)$.  We use the boundary condition of $\rho_\star
\sigma^2 \rightarrow 0$ at the outer radius of the cluster, which we
take to be 17 pc as described above.

To determine $\rho_\star(r)$, we deproject the light distribution of
the cluster via an Abel transformation.  We use the results of the
circularly symmetric GALFIT decomposition of the WFPC2 image as the
model for the observed light distribution, assuming that the four
narrowest GALFIT Gaussian components belong to the cluster.  These
Gaussian components are individually deprojected by Abel
transformations, and the resulting $\rho_\star(r)$ is the sum of these
four deprojected functions.  

To compare to the measured velocity dispersion of the cluster we
integrate along the line of sight through the cluster:
\begin{equation} 
\sigma^{2}(R) = \frac{2}{I_\star(R)} \int_{R}^{r_\mathrm{out}} \left ( 1 -
\beta \frac{R^{2}}{r^2} \right ) \frac{\rho_{\star} \sigma_{r}^{2}
r}{\sqrt{r^2-R^2}} dr,
\label{eq:sigmaLOS}
\end{equation} 
where $R$ is the projected radius on the plane of the sky and
$I_\star(R)$ is the surface brightness at $R$.  We carry out the
integration up to a radius of $r_\mathrm{out}=17$ pc, since this is
the assumed outer radius of the cluster.  Given the multi-component
light distribution determined from GALFIT, equations~\ref{eq:jeans}
and~\ref{eq:sigmaLOS} cannot be solved analytically, so to obtain
$\sigma^{2}$ we first numerically integrate the Jeans equation to
obtain $\sigma_r^2$, and then integrate this solution along the line
of sight using equation~\ref{eq:sigmaLOS} to obtain $\sigma^2$. 

Finally, to obtain the integrated line-of-sight velocity dispersion
for comparison with the observed value, we average the line-of-sight
velocity dispersion over the cluster, weighted by surface brightness:
\begin{equation} 
\langle \sigma^2 \rangle = \frac{2\pi \int \sigma^2(R) I_\star(R)
RdR}{2\pi \int I_\star(R) RdR} .
\label{eq:avg}
\end{equation} 

In the form above, the Jeans equation assumes spherical symmetry;
given the small observed ellipticity of the light distribution of the
cluster, as well as the fact that the circularly symmetric GALFIT
model fits the cluster's radial profile well overall, this provides a
good approximation to the dynamical mass. Further, this approach
assumes that the cluster is non-rotating. Though the data do not give
us any information on the rotation of this cluster, the small
ellipticity can be interpreted as a small amount of flattening of the
cluster due to rotation.

We model the mass distribution, $M(r)$, in terms of a component
proportional to the stellar luminosity profile (assuming a spatially
uniform stellar mass-to-light ratio) and a component representing the
central point mass. This procedure allows us to examine a
two-dimensional parameter space of the stellar mass-to-light ratio,
$M/L$, and the central point mass, \mbh.  As noted by \citet{bok99},
an upper limit to \mbh\ can be determined by modeling the cluster in
the limit that the stellar $M/L$ approaches zero.  

Since the cluster's mass and radius imply a half-mass relaxation time
of order $\sim5$ Gyr \citep{mer08}, and the cluster probably contains
significant population components younger than this, the cluster may
not be relaxed and it could have an anisotropic velocity dispersion.
When scanning the parameter space, we fix the velocity anisotropy and
assume it to be constant throughout the cluster.  To explore the
possible effects of anisotropy, we calculate models for three fixed
values of the $\beta$ parameter, $\beta = -1,0$, and 0.5. The values
$\beta = -1$ and 0.5 correspond to ratios of tangential-to-radial
velocity dispersions of $\sqrt{2}:1$ and $1:\sqrt{2}$, respectively.

\begin{figure}
\plotone{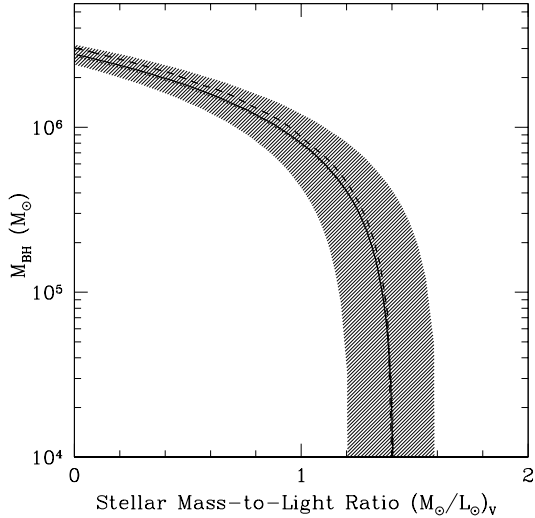}
\caption{Dynamical modeling results.  The curves show the locus of
  models yielding a line-of-sight velocity dispersion of $\sigma=43$
  \kms\ in the stellar $(M/L)_V$ vs.\ \mbh\ plane. The dotted curves
  are for radially anisotropic models, the solid curves are for the
  isotropic models, and the dashed curves are for the tangentially
  isotropic models. The shaded band represents the $1\sigma$
  uncertainty range on \sigmastar\ of 40-46 km s$^{-1}$ for the
  isotropic model.
\label{MLvsMBH}}
\end{figure}

In Figure~\ref{MLvsMBH}, we show the resulting two-dimensional
parameter space of \mbh\ vs.\ stellar mass-to-light ratio. For the
specific case of isotropic velocity dispersion, the shaded band in
Figure~\ref{MLvsMBH} shows the projected contours of line-of-sight
velocity dispersions corresponding to the $1\sigma$ error range on our
measurement of the cluster velocity dispersion of $\sigma = 43\pm3$
\kms. We see from Figure~\ref{MLvsMBH} that, in the limit that the
stellar mass-to-light ratio approaches zero, we obtain a strict upper
limit to the mass of the black hole, $3 \times 10^6$ \msun.  Since we
are averaging the velocity dispersion over the entire surface area of
the cluster as in equation~\ref{eq:avg}, this result turns out to be
nearly independent of velocity anisotropy.  We also find that, in the
limit that $\mbh \rightarrow 0$, the stellar mass-to-light ratio is
$1.4\pm0.2 ~(\msun/L_{\odot,V})$, again independent of the velocity
anisotropy.

\begin{figure}
\plotone{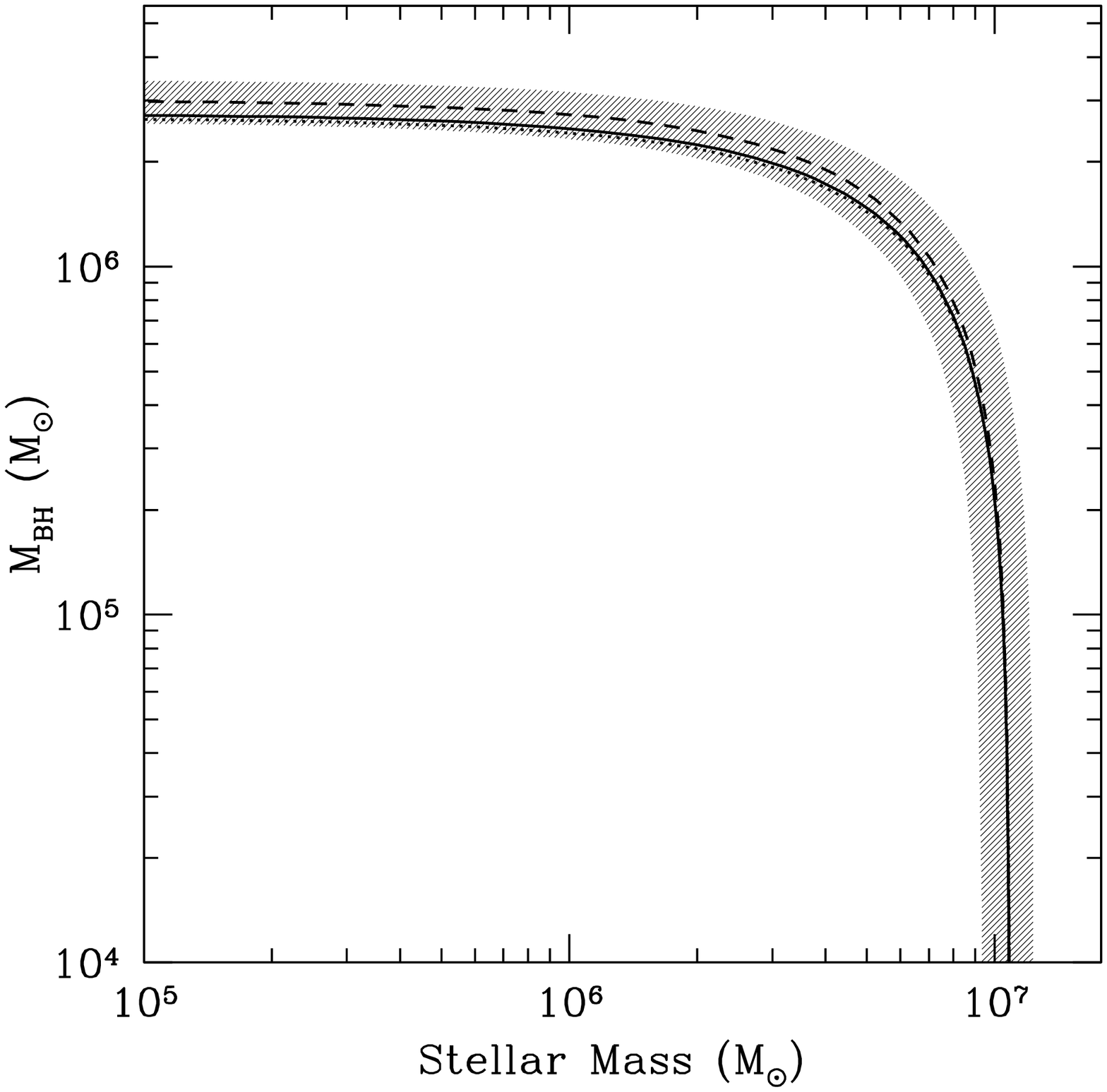}
\caption{Dynamical modeling results.  The curves show the locus of
  models yielding a line-of-sight velocity dispersion of $\sigma=43$
  \kms\ in the stellar mass vs.\ \mbh\ plane.  Similar to
  Figure~\ref{MLvsMBH}, the dotted curves are for radially anisotropic
  models, the solid curves are for the isotropic models, and the
  dashed curves are for the tangentially isotropic models. The shaded
  band represents the $1\sigma$ uncertainty range on \sigmastar\ of
  40-46 km s$^{-1}$ for the isotropic model.
\label{MassvsMBH}}
\end{figure}

In Figure~\ref{MassvsMBH}, we show the two-dimensional parameter space
of black hole mass and integrated stellar mass within 17 pc. In the
limit that $M_{BH} \rightarrow 0$, the stellar mass within 17 pc is
constrained to be within $(0.95-1.2) \times 10^7$ \msun.  This derived
total stellar mass is on the larger end of the distribution of those
measured by \cite{wal05}, due to the relatively high stellar velocity
dispersion measured for NGC 3621.  We also note that this result
implies a value of $f\approx6$ for the numerical factor in the virial
theorem mass estimate (Equation \ref{eq:virial}).

We also computed a set of models in which the outermost Gaussian
component from the WFPC2 GALFIT decomposition (with FWHM = 3\farcs1 or
99.2 pc) was assumed to belong to the cluster rather than to the
galaxy disk, to assess the impact that this choice has on the model
results.  The difference proved to be insignificant, because the width
of this component is so large that only a small fraction of its mass
lies within our fiducial 17 pc radius for the dynamical mass
calculation.  In the limit of $\mbh=0$, the dynamical mass for the
cluster changed by $<1\%$ when the outermost Gaussian component was
included as part of the cluster.  Similarly, the upper limit to \mbh\
is unchanged if this component is added to the cluster model, and the
error contours in Figures \ref{MLvsMBH} and \ref{MassvsMBH} are
essentially unaffected by this change.

It bears repeating that in our mass estimates we assume that there is
no rotational component contributing to the measured velocity
dispersion. As argued above, this is a plausible assumption, although
it is at this time not possible to determine what fraction of the
observed velocity dispersion comes from ordered rotational motion.
Any rotational motion will affect the above mass estimates, though
determining by how much requires a full rotational model of the
cluster, which depends not only on the rotational speed but also the
inclination and eccentricity of the cluster.  If the cluster were
rapidly rotating \emph{and} oriented with its rotation axis close to
our line of sight, then our method would underestimate the true mass
of the cluster and underestimate the upper limit to \mbh.  With the
data currently available, it is not possible to resolve the internal
kinematics of the cluster or to constrain the cluster's rotation, or
to measure the rotation speed or dispersion of the inner disk in the
neighborhood of the cluster.  However, it is unlikely that the cluster
would be rotating in a nearly face-on orientation, since the host
galaxy as a whole has an inclination of $i=65.6\arcdeg$ \citep{pat03}.
The high-resolution kinematic study of the nuclear cluster in the
edge-on galaxy NGC 4244 by \citet{seth08b} demonstrates that rotation
can be dynamically important in nuclear clusters, and similar
observations of the NGC 3621 cluster will be important for refining
the cluster mass estimate.

\citet{mer87} has shown that a lower limit to the mass of a cluster
can be derived from the virial theorem, based on the extreme limiting
case in which all of the mass is contained at the central point of the
cluster.  This minimum mass is given by $G M_\mathrm{min} = 3
\sigmastar^2/\langle r^{-1} \rangle$, where $1/\langle r^{-1} \rangle$
is the harmonic mean radius of stars in the cluster and \sigmastar\ is
the line-of-sight velocity dispersion. Averaged over the
three-dimensional, de-projected light distribution, $\langle r^{-1}
\rangle = 0.5$ pc\per, and the resulting lower limit to the cluster's
dynamical mass is $\sim 2.6 \times 10^6$ M$_\odot$.  This limit
corresponds to our Jeans equation model for a cluster in which all of
the mass is in the central point source, and the masses derived by
these two methods turn out to be nearly identical.  We interpret this
in the following way.  The minimum possible total mass of the cluster
is found for the limit in which all of the mass is in a central point,
whether calculated by the virial theorem or by more detailed Jeans
equation modeling.  As our Jeans equation modeling demonstrates, the
lower limit to the mass of the cluster overall is the same as the
upper limit to the mass of a black hole in the cluster.  Therefore,
the quantity $3 \sigma^2/G\langle r^{-1} \rangle$ may be useful as a
simple virial estimator of the maximum possible mass of a black hole
in a star cluster.  This could prove useful as a shortcut for deriving
upper limits to black hole masses in larger surveys of nuclear
clusters, and the limit derived in this way is substantially smaller
than the total cluster mass derived from Equation \ref{eq:virial}.

In principle, the upper limit to \mbh\ could be lowered further if the
stellar mass-to-light ratio could be constrained more accurately from
multiband photometry and spectroscopy.  From
\S\ref{sectionstellarmass}, the photometric data implies a $V$-band
stellar mass-to-light ratio in the range $(1.4-2.0)
\msun/L_{\odot,V}$.  If we conservatively assume that $M/L_V>1$ then
the upper limit to \mbh\ would be $\sim10^6$ \msun.  Since our stellar
mass estimates from photometry are very uncertain, we keep the extreme
case of $\mbh<3\times10^6$ as our best estimate of the upper limit to
the black hole mass in this galaxy.

\section{Discussion}

\subsection{AGN Energetics}

The AGN in NGC 3621 is faint and may be heavily obscured, and it is
difficult to determine its bolometric luminosity.  \citet{sat07} found
a correlation between mid-infrared [\ion{Ne}{5}] emission-line
luminosity and bolometric luminosity for a small sample of nearby AGNs
with well-sampled spectral energy distributions, and used this
correlation to derive an estimate of \lbol\ for NGC 3621 based on its
observed [\ion{Ne}{5}] flux.  They found $\lbol \approx
5\times10^{41}$ ergs s\per, which is a substantial luminosity for an
AGN in a late-type spiral. For comparison, NGC 4395 has $\lbol\approx
5\times10^{40}$ ergs s\per\ and $\lbol/\ledd \approx 10^{-3}$
\citep{mor99, pet05}.  If the infrared-based estimate of \lbol\ in NGC
3621 is correct, then the AGN must be almost entirely obscured in the
optical; this would be consistent with its Type 2 classification and
starlight-dominated continuum.  Combining the value of \lbol\ from
\citet{sat07} with our constraints on the black hole mass, we have
$\lbol/\ledd \gtrsim 10^{-3}$.

The [\ion{O}{3}] $\lambda5007$ flux measured from our ESI spectrum is
$(22.4\pm0.4)\times10^{-16}$ ergs s\per\ after correcting for Galactic
extinction.  The uncertainty reflects only the statistical error in
the extracted spectrum; additional error will arise from the unknown
difference in slit losses between the galaxy and standard star
observation.  As described above in \S\ref{section:emissionlines}, the
ESI spectrum probably includes only $\sim10\%$ of the total
[\ion{O}{3}] flux from the AGN, so our measurement gives a lower limit
to the true emission-line luminosity.

\citet{hec04} find that for luminous Seyfert 1 galaxies,
$\lbol/L$([\ion{O}{3}])$\approx 3500$ with a scatter of approximately
0.38 dex.  Applying this [\ion{O}{3}] bolometric correction to the
lower limit for the [\ion{O}{3}] luminosity of the NGC 3621 nucleus,
we arrive at a lower limit of $\lbol > 4\times10^{40}$ ergs s\per\
based on the optical spectrum.  This is an order of magnitude lower
than the estimated \lbol\ from the \emph{Spitzer} spectrum.  If the
ESI spectrum represents 10\% of the total nuclear [\ion{O}{3}] flux,
then the full [\ion{O}{3}] luminosity would give an estimate of \lbol\
that is essentially consistent with the luminosity derived from the
[\ion{Ne}{5}] emission by \citet{sat07}.  Ultimately, the best
constraints on the AGN energy budget and obscuration could be obtained
from new X-ray observations, but there are none available at present.
Perhaps most important, detection of a compact, hard X-ray source in
the center of NGC 3621 would provide confirmation that it is genuinely
an AGN.

\subsection{Black Holes in Late-Type Spirals}

Our primary result is that NGC 3621, which is a bulgeless or nearly
bulgeless disk galaxy, contains a nuclear star cluster in which the
maximum possible mass of a central black hole is $3\times10^6$ \msun.
NGC 3621 is one of only a few examples of a bulgeless disk galaxy
containing both a nuclear star cluster and a spectroscopically
detected AGN.  Under the standard assumption that the nuclear activity
is driven by gas accretion onto a black hole, our results provide new
evidence that some bulgeless spiral galaxies do contain low-mass
central black holes.  It is important to acknowledge that the
detection of nuclear activity in NGC 3621 remains a good deal less
secure than, for example, the well-studied Seyfert 1 nucleus in the
late-type spiral galaxy NGC 4395.  Still, the high-ionization
[\ion{Ne}{5}] emission found by \citet{sat07} and the Seyfert-type
optical spectrum both provide a reasonable basis for classifying this
object as an AGN.  In this case, the low luminosity of the AGN is
actually somewhat advantageous: with no bright nuclear point source it
is possible to obtain accurate measurements of both the stellar
velocity dispersion and the structural properties of the nuclear star
cluster.  Such measurements would be far more difficult for a galaxy
like NGC 4395, which contains a bright, unobscured AGN within its
nuclear cluster.

\citet{ros06} found an anticorrelation between morphological $T$-type
and nuclear cluster mass, and a very weak correlation (with very large
scatter) between total host galaxy $B$-band luminosity and nuclear
cluster mass.  Our dynamical models imply a stellar mass of $\sim10^7$
\msun\ for the NGC 3621 nuclear cluster if the black hole mass is
small compared with the cluster's total mass.  NGC 3621 has
$T=6.9\pm0.4$ and $\log(L_B/L_{\odot,B}) = 9.9$ \citep{pat03}.  With
these values, and assuming a stellar mass of $10^7$ for the NGC 3621
nuclear cluster, it is a fairly typical object in comparison with the
\citet{ros06} sample, falling within the scatter of both the cluster
mass vs.\ $T$-type and cluster mass vs.\ host luminosity
relationships.  The only property of this cluster that appears unusual
relative to previously studied nuclear clusters in late-type spirals
is its larger velocity dispersion.  

\begin{figure}
\plotone{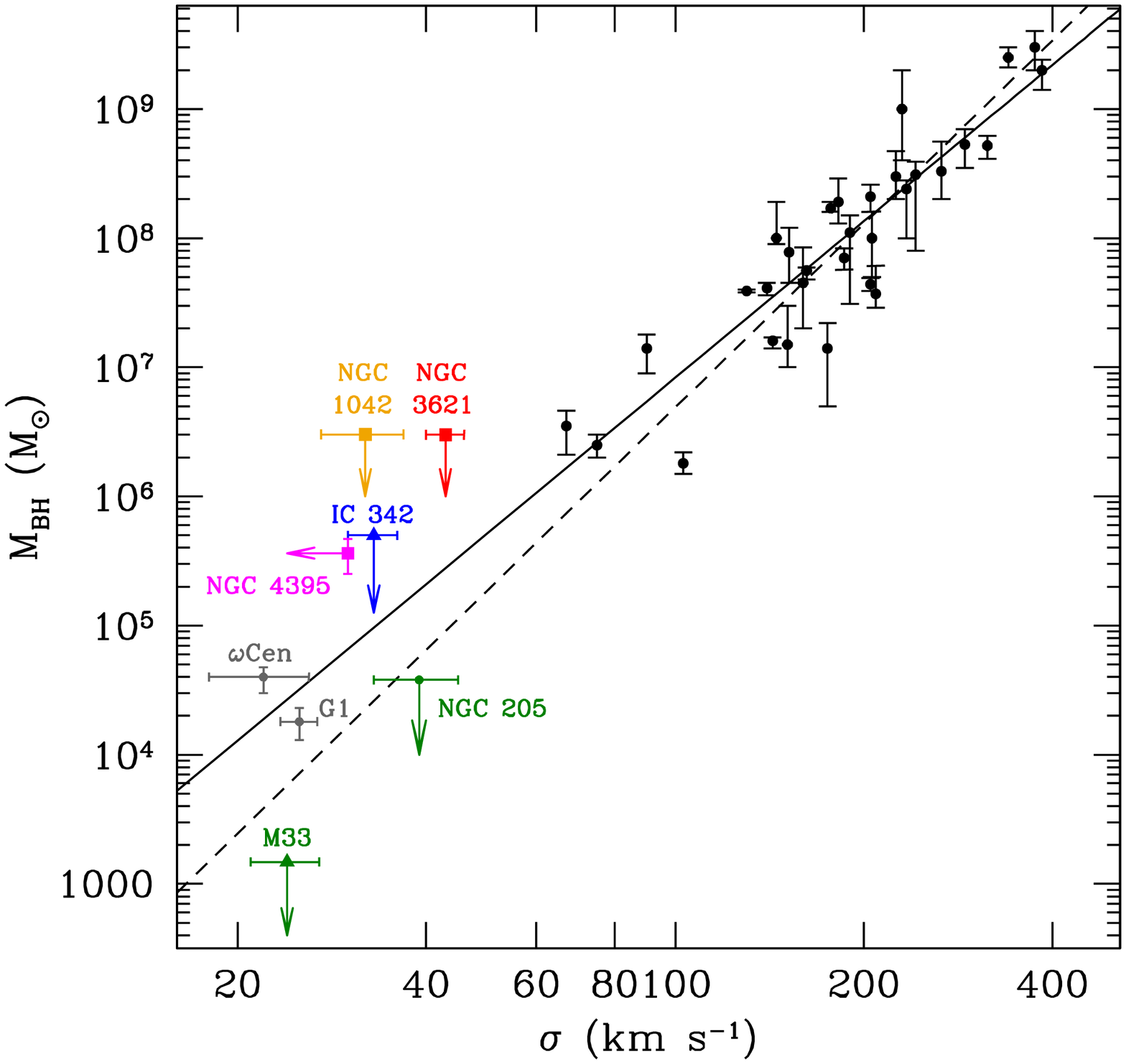}
\caption{The \msigma\ relation including nuclear clusters in late-type
  spiral galaxies.  Black circles are objects from the compilation of
  \citet{tre02}.  Filled squares represent nuclear clusters containing
  AGNs, including NGC 3621 (this work), NGC 4395 \citep{fh03,pet05},
  and NGC 1042 \citep{wal05}.  Triangles represent nuclear clusters in
  late-type spirals without AGNs: IC 342 \citep{bok99}, M33
  \citep{geb01}.  Globular clusters with stellar-dynamical detections
  of black holes are G1 \citep{geb05} and $\omega$ Cen \citep{noy08}.
  The upper limit to the black hole mass in NGC 205 is from
  \citet{val05}.  Lines denote fits to the \msigma\ relation from
  \citet{tre02} (solid line) and \citet{mf01} (dashed line).
\label{msplot}}
\end{figure}

The location of NGC 3621 in the \msigma\ relation is shown in Figure
\ref{msplot}.  In this case, we use the sample compiled by
\citet{tre02} for comparison, and include measurements or constraints
on \mbh\ and \sigmastar\ for other late-type spiral galaxies (NGC
4395, IC 342, M33, and NGC 1042), for the dwarf elliptical NGC 205
\citep{val05}, and for the globular clusters G1 and $\omega$ Cen
\citep{geb05,noy08}.  For the late-type spirals, \sigmastar\ is taken
to be the integrated line-of-sight velocity dispersion of the nuclear
star cluster.  The upper limit for NGC 3621 falls well above the
\msigma\ relation extrapolated to low $\sigma$.  Some studies of the
low-mass end of the \msigma\ relation have suggested a possible
flattening in slope at the low-mass end \citep{bgh05,gh06b,wy06}, and
it is important to test these claims with new dynamical measurements
of black hole masses in low-dispersion galaxies, but our results do
not provide any strong new constraints on the \msigma\ slope.

The upper limit to \mbh\ in NGC 1042 also lies considerably above the
\msigma\ relation, but we note that this upper limit (from Shields
\etal\ 2008) is simply the total dynamical mass of the cluster from
Walcher \etal\ (2005, 2006), which is dominated by the stars in the
cluster.  The upper limit could be considerably tightened by
calculating dynamical models with a central point mass in addition to
an extended stellar component, as described above.  Similarly, it
would be interesting to compute upper limits based on this technique
for all of the objects having stellar velocity dispersion measurements
from \citet{wal05}, even for nuclear clusters that do not contain
AGNs, since the more stringent upper limits that could be derived from
this analysis could provide important new constraints on black hole
demographics in late-type spirals.  

With just a few nearby examples of very late-type spirals that are
known to host AGNs, and just one object (M33) in which a massive black
hole can be essentially ruled out, we have only a very incomplete
picture of black hole demographics in late-type spirals.  An important
motivation for improving the black hole census in low-mass galaxies
with small velocity dispersions is that the \msigma\ relation and
black hole occupation fraction together encode key information about
the mass scale of black hole seeds and the efficiency of seed
formation, as discussed by \citet{vol08}.  Scenarios in which black
holes begin as massive ($\sim10^5$ \msun) seeds formed by collapse of
low angular momentum gas at high redshift \citep{kbd04,ln06} generally
lead to low black hole occupation fractions in galaxies with small
velocity dispersions.  On the other hand, if black hole seeds are
formed from the remnants of Population III stars and have masses of
$\sim100$ \msun, then the merger tree calculations described by
\citet{vol08} predict high black hole occupation fractions in low-mass
galaxies, but a steep slope for the lower end of the \msigma\
relation, with some galaxies hosting very tiny ($<10^4$ \msun) black
holes.  While these models still do not incorporate all of the
relevant physical processes involved in black hole growth (such as
gravity-wave recoil kicks), it is becoming clear that a full census of
black holes in dwarf and late-type galaxies can be a fundamentally
important diagnostic of the mass scale of black hole seeds at high
redshift.  Much progress can be made through a combination of
multiwavelength AGN surveys \citep[e.g.,][]{gh07, gal08, seth08a,
sat08, des08} and improved stellar-dynamical searches for low-mass
black holes, although a full census of the black hole content in
low-mass galaxies is very far from being feasible at present.

Another possibility is that black holes in galaxies such as NGC 3621
might have formed at a later evolutionary stage, within the nuclear
clusters themselves. The high densities of nuclear star clusters, as
well as the likelihood of ongoing gas inflow from the host galaxy disk
\citep{sbm03,mil04}, offer suggestions that nuclear clusters could be
favorable environments for the formation and early growth of massive
black holes.  In sufficiently dense young star clusters with short
dynamical friction timescales, intermediate-mass black holes could be
formed by runaway collisions of massive stars in the cluster core
\citep{por04}.  This process is, however, unlikely to operate within
typical nuclear star clusters in disk galaxies.  As noted by
\citet{wal05}, the estimated dynamical friction timescale for a 100
\msun\ star to sink to the center of a nuclear cluster is longer than
the star's lifetime, and as a result the runaway process described by
\citet{por04} would not occur.  The same is true of NGC 3621; from
Equation 1 of \citet{por04} we estimate a dynamical friction timescale
of $\sim50$ Myr for a 100 \msun\ star to sink to the cluster center,
which is much longer than the lifetime of such a massive star.  Still,
more extreme conditions might occur during the early history of
nuclear cluster formation, and the question of whether young nuclear
star clusters might be favorable sites for black hole formation
deserves further theoretical consideration.

\section{Summary and Conclusions}

Our main conclusions are as follows.

\begin{enumerate}

\item NGC 3621 has a Seyfert 2 optical spectrum, consistent with the
  AGN identification based on its mid-infrared spectrum by
  \citet{sat07}.  This adds further support for the hypothesis that a
  black hole is present.

\item For the nuclear star cluster, we find an effective radius of
  $r_e=4.1$ pc, a line-of-sight stellar velocity dispersion of
  $\sigmastar=43\pm3$ \kms, and an absolute magnitude of $M_V = -12.3$
  mag (uncorrected for internal extinction).  Simple estimates based
  on the available \hst\ photometry suggest a stellar mass of
  $(1-3)\times10^7$ \msun.

\item From two-dimensional image decomposition of 2MASS images, we
  find that the galaxy is dominated by an exponential disk, with at
  most a very faint possible pseudobulge component being present.  We
  derive an upper limit of $M_K = -17.6$ mag to the luminosity of any
  bulge or pseudobulge component.

\item From stellar-dynamical modeling of the nuclear cluster, we find
  an upper limit to the black hole mass of $\mbh < 3\times10^6$.  This
  result is insensitive to possible velocity anisotropy within the
  cluster.  The upper limit is based on the extreme assumption of
  $M/L=0$ for the stellar population.  With improved photometric and
  spectroscopic observations it would be possible to better constrain
  the stellar mass-to-light ratio and reduce the upper limit on \mbh.
  Assuming that the cluster's black hole mass is small in comparison
  with its stellar mass, the dynamical models give a stellar mass of
  $\sim1\times10^7$ \msun\ for the cluster.

\end{enumerate}

These results add to the emerging body of evidence that some
late-type, bulgeless galaxies do contain low-mass central black holes,
and that black holes can be found inside some nuclear star clusters.
It remains the case, however, that there are only a few late-type
spirals in which a the presence of a massive black hole has been
either confirmed (based on nuclear activity) or ruled out (from
dynamical modeling).

Further observations of the NGC 3621 nuclear cluster will lead to a
more complete picture of the properties of this object.  The most
pressing observational need is for X-ray observations, to provide more
definitive confirmation of the AGN interpretation and a better
estimate of the AGN luminosity.  Improved measurements of the stellar
content of the nuclear cluster can be obtained from high-resolution,
high-S/N spectra extending farther to the blue than our ESI data, in
addition to improved multi-band \hst\ imaging covering the
near-ultraviolet to near-infrared.  Future extremely large
ground-based telescopes with adaptive optics may make it possible to
spatially resolve the kinematic substructure of the nuclear cluster,
and this could lead to dramatically improved constraints on the black
hole mass.

\acknowledgments

Research by A.J.B. and M.C.B. is supported by NSF grant AST-0548198,
and research by L.E.S. is supported by NSF grant AST-0607746.  We
thank Rachel Kuzio de Naray for helpful conversations, and an
anonymous referee for suggestions that improved this work.  Data
presented herein were obtained at the W.M. Keck Observatory, which is
operated as a scientific partnership among Caltech, the University of
California, and NASA. The Observatory was made possible by the
generous financial support of the W.M. Keck Foundation.  The authors
wish to recognize and acknowledge the very significant cultural role
and reverence that the summit of Mauna Kea has always had within the
indigenous Hawaiian community.  We are most fortunate to have the
opportunity to conduct observations from this mountain.  This
publication is based on observations made with the NASA/ESA Hubble
Space Telescope, obtained from the Data Archive at the Space Telescope
Science Institute, which is operated by the Association of
Universities for Research in Astronomy, Inc., under NASA contract NAS
5-26555. These observations are associated with programs 5446, 9492,
and 11080.  This publication makes use of data products from the Two
Micron All Sky Survey, which is a joint project of the University of
Massachusetts and the Infrared Processing and Analysis
Center/California Institute of Technology, funded by the National
Aeronautics and Space Administration and the National Science
Foundation.  This research has made use of the NASA/IPAC Extragalactic
Database (NED) which is operated by the Jet Propulsion Laboratory,
California Institute of Technology, under contract with the National
Aeronautics and Space Administration.

\end{document}